\documentclass{svproc}
\usepackage{times}
\usepackage{helvet}
\usepackage{courier}
\usepackage[hyphens]{url}
\usepackage{graphicx}
\usepackage{todonotes}
\usepackage{float}
\usepackage{tabularx} 
\usepackage{booktabs} 
\usepackage{adjustbox} 
\usepackage{amsmath}
\usepackage{rotating}
\usepackage{caption}
\begin{document}
\mainmatter              
\title{Judging the Judges: Human Validation of Multi-LLM Evaluation for High-Quality K--12 Science Instructional Materials}
\titlerunning{Hamiltonian Mechanics}  
%

\author{
Peng He\inst{1} \and
Zhaohui Li\inst{2} \and
Zeyuan Wang\inst{1} \and
Jinjun Xiong\inst{2} \and
Tingting Li\inst{1}
}

\authorrunning{He et al.}

\tocauthor{Peng He, Zhaohui Li, Zeyuan Wang, Jinjun Xiong, Tingting Li}

\institute{
Washington State University, Pullman, WA, USA\\
\email{tingting.li1@wsu.edu}
\and
University at Buffalo, State University of New York, Buffalo, NY, USA\\
}

\maketitle              

\begin{abstract}
Designing high-quality, standards-aligned instructional materials for K--12 science is time-consuming and expertise-intensive. This study examines what human experts notice when reviewing AI-generated evaluations of such materials, aiming to translate their insights into design principles for a future GenAI-based instructional material design agent. We intentionally selected 12 high-quality curriculum units across life, physical, and earth sciences from validated programs such as OpenSciEd and Multiple Literacies in Project-Based Learning. Using the EQuIP rubric with 9 evaluation items, we prompted GPT-4o, Claude, and Gemini to produce numerical ratings and written rationales for each unit, generating 648 evaluation outputs. Two science education experts independently reviewed all outputs, marking agreement (1) or disagreement (0) for both scores and rationales, and offering qualitative reflections on AI reasoning. This process surfaces patterns in where LLM judgments align with or diverge from expert perspectives, revealing reasoning strengths, gaps, and contextual nuances. These insights will directly inform the development of a domain-specific GenAI agent to support the design of high-quality instructional materials in K--12 science education.
\end{abstract}
\section{Introduction}

Science education in the United States has been shaped by a strong consensus around
\textit{A Framework for K–12 Science Education}~\cite{NRC2012} and the \textit{Next Generation Science
Standards (NGSS)}~\cite{NGSS2013}. Over 40 states have adopted or adapted
these standards, reflecting a national vision for rigorous and equitable science learning.
However, realizing this vision in classrooms requires high-quality instructional materials
and sustained teacher professional learning~\cite{Achieve2016a,Achieve2016b,BSCS2019,Carnegie2017,NASEM2018}. Instructional materials translate policy
into practice: they provide teachers with concrete resources, embed disciplinary
standards, and must address equity and diversity\cite{campbell2021instructional}. Developing such materials is a
scholarly and practice-based challenge—one that motivates the present study on how
human experts and AI can contribute to evaluating and improving K–12 science
curriculum resources.

High-quality instructional materials are central to achieving equitable, standards-aligned science education \cite{campbell2021instructional}. While Generative AI (GenAI) tools have shown promise in automating curriculum evaluation, their judgments often lack validation against the nuanced insights of human experts \cite{Yuan2024CourseEvaluationLLM,AbdulWahid2025CurriculumAlignedMCQ,Clark2025AutoEvaluation}. This work addresses this gap by systematically examining how experts evaluate AI evaluations and by translating these insights into design principles for future AI agents that support instructional material development.

While prior studies have explored the use of large language models for rubric-based
evaluation of instructional materials or examined agreement between AI and human
raters, most existing work focuses on single-model performance or treats human
validation primarily as a benchmarking step. In contrast, this study adopts a
comparative, multi-LLM perspective and foregrounds expert review as an analytic
lens rather than a correctness check. By systematically examining where and why
expert judgments align with or diverge from multiple LLMs across rubric dimensions,
our approach reveals epistemic patterns in AI reasoning that are often obscured by
aggregate accuracy metrics. This enables the translation of expert insights into
actionable design principles for disagreement-aware, context-sensitive GenAI agents,
advancing AI-based curriculum evaluation beyond automated scoring.

\section{Related Work}
We review two key strands of research: (1) AI-assisted evaluation of K--12 instructional materials, and (2) comparative studies of human and AI evaluative judgments. While prior work has demonstrated that LLMs can produce rubric-based evaluations, little is known about how domain experts perceive, validate, or contest these AI outputs.
\subsubsection{Evaluating K--12 Science Instructional Materials}
High-quality curriculum evaluation has been studied extensively in science education. The Educators Evaluating the Quality of Instructional Products (EQuIP) rubric developed by Achieve provides structured criteria for assessing alignment to the Next Generation Science Standards (NGSS) \cite{achieve2016equip}. Research by Penuel et al. has examined the design and implementation of NGSS-aligned materials and the practical challenges teachers face in using them effectively \cite{penuel2017rpp}. Studies have highlighted difficulties in applying these evaluation frameworks reliably across settings \cite{nordine2021promoting}. However, most of this work has focused on human-led processes and has not yet explored the integration of AI systems for curriculum evaluation

\subsubsection{Large Language Models for Educational Assessment}
Large Language Models (LLMs) have shown promise in automating diverse educational assessment tasks. Recent research has demonstrated that LLMs can generate plausible rubric-based evaluations of student writing and provide formative feedback \cite{li2024culturally,hackl2023gpt,tate2024can,huang2024application,matsukawa2024development,bernabei2023students}. For example, Hackl (2023) found that GPT‑4 exhibited high internal consistency when used to evaluate educational assessments, with intra-class correlation scores ranging from 0.94 to 0.99 across repeated trials \cite{hackl2023gpt}. Tate et al. (2024) reported that ChatGPT showed moderate agreement with human raters in holistic essay scoring, with AI ratings falling within one point of human scores in the majority of cases \cite{tate2024can}. Yet concerns remain about hallucinations, bias, and alignment with human judgment, particularly in high-stakes settings \cite{li2024culturally,ganguli2023capacity}. Moreover, few studies have compared how multiple LLMs perform on the same educational evaluation tasks within K--12 science contexts.

\vspace{-5pt}
\subsubsection{Human Validation of AI-Generated Evaluations}

Human-in-the-loop validation is essential to ensure that AI-generated evaluations are both accurate and pedagogically meaningful. Prior work in writing assessment has shown that expert review can expose discrepancies in AI scoring and identify areas for rubric refinement\cite{crossley2013using}. Similarly, studies in science education emphasize the need for transparency and traceability in AI-generated feedback used in classroom contexts \cite{Li2024CognitiveSynergy}. Beyond accuracy, such validation helps surface reasoning strategies and contextual nuances often invisible to quantitative metrics. This aligns with broader calls for co-design approaches in human–AI systems, where teachers actively shape the role of AI in classroom decision-making\cite{holstein2019co}.
\vspace{-5pt}
\subsubsection{Designing AI Agents for Curriculum Development}
There is growing interest in leveraging generative AI to support curriculum design and instructional planning \cite{Dickey2023GAIDE,Jaramillo2024AIDriven,Tavakoli2021Hybrid}. In line with this, the U.S. Department of Education recently issued guidance encouraging the use of federal grant funds to integrate AI—particularly for producing high-quality instructional materials, enhancing tutoring, and supporting college and career counseling—while emphasizing compliance with regulatory frameworks and stakeholder engagement \cite{ED2025DCL}. This federal endorsement highlights the urgency and legitimacy of research aimed at designing responsible, context-sensitive AI agents for education. Research indicates that using general-purpose models without domain-specific alignment may result in inconsistent pedagogical value and limited contextual adaptation \cite{zawacki2019systematic}. Similarly, concerns have been raised about the lack of integration between AI evaluations and educator heuristics \cite{holstein2019co}. Our study contributes by designing GenAI evaluation agents whose outputs are validated by human experts—particularly focusing on NGSS-aligned K–12 science instructional materials to ensure both fidelity to standards and instructional relevance.

\section*{Research Questions}

\noindent
\textbf{RQ1:} To what extent do human experts agree with the numerical ratings and written rationales generated by different LLMs when applying the EQuIP rubric to evaluate high-quality K–12 science instructional materials?

\vspace{0.5em}

\noindent
\textbf{RQ2:} What reasoning strengths, weaknesses, and contextual considerations do experts identify when reviewing LLM-generated evaluations, and how do these vary by model and rubric dimension?



\begin{table*}[t]
\centering
\small

\caption{System and User prompts used in all evaluations.}
\label{tab:prompts}
\begin{tabular}{p{0.18\linewidth} p{0.75\linewidth}}
\hline
\textbf{System Prompt} & 
You are a professional school teacher; you need to evaluate the learning activity plan for the Physical Science lesson based on the attached \textit{Criteria for Designing and Evaluating NGSS 3D Learning Activities}. \\
\hline
\textbf{User Prompt} & 
Please read through the attached learning activity and provide a table that evaluates each Category. The evaluation needs to include:  
(1) Evidence of Quality (3: Extensive, 2: Adequate, 1: Inadequate, 0: None)  
(2) Specific evidence from materials (what happened / where did it happen) and reviewer’s reasoning (how / why is this evidence)  
(3) Suggestions for improvement.  

Don't forget all Criteria:  
Category I: NGSS 3D Design A. Explaining Phenomena/Designing Solution;  
Category I: NGSS 3D Design B. Three Dimensions;  
Category II: NGSS Instructional Supports A. Relevance and Authenticity;  
Category II: NGSS Instructional Supports B. Student Ideas;  
Category II: NGSS Instructional Supports C. Scientific Accuracy;  
Category III: Monitoring NGSS Student Progress A. Monitoring 3D student performances;  
Category III: Monitoring NGSS Student Progress B. Feedback;  
Category III: Monitoring NGSS Student Progress C. Unbiased tasks/items. \\
\hline
\end{tabular}

\end{table*}

\vspace{0.5em}
\section{Methods}
\subsection{Data Collection}

We selected 12 curriculum units covering life, physical, and Earth sciences, drawn from validated, high-quality programs—including OpenSciEd, Multiple Literacies in Project-Based Learning (ML-PBL), and other NGSS-aligned units recognized with the NGSS Quality Badge. Each unit was evaluated using the 9-item EQuIP rubric (see Table~\ref{tab:model-human-example}) by three leading large language models—GPT-4o, Claude, and Gemini—via engineered prompts (see Table~\ref{tab:prompts}). Each model generated both numerical scores and written rationales. Two science education experts independently reviewed each model output, indicating agreement (1) or disagreement (0) with the assigned score and rationale, resulting in a total of 648 expert judgments.

\subsection{Data Analysis}

To address \textbf{RQ1}, we assessed alignment between large language models (LLMs) and human raters by computing inter-model and model–human agreement. For each model (GPT-4o, Claude, Gemini), we calculated score-level agreement using Cohen’s $\kappa$ for pairwise comparisons \cite{cohen1960coefficient} and Fleiss’ $\kappa$ for multi-rater summary \cite{fleiss1971measuring}. We also reported raw percent agreement, following guidance from \cite{landis1977measurement}, given the practical importance of absolute agreement in instructional contexts. Chi-square tests were used to detect significant differences across science domains (Physical, Life, Earth) and rubric dimensions (e.g., DCI integration, use of evidence).

To address \textbf{RQ2}, we conducted thematic analysis of expert comments on LLM-generated evaluations. Using grounded theory techniques \cite{strauss1990basics}, we coded themes related to reasoning quality, omissions, generalizations, and pedagogical insight. We then cross-tabulated codes by model and rubric dimension to identify systematic differences in evaluative stance. Finally, we analyzed the models’ written justifications alongside expert feedback to surface underlying epistemic assumptions.

\subsubsection{Anticipated Outcomes}
We expect to (1) benchmark model-specific agreement rates with experts, 
(2) map reasoning strengths and weaknesses, and 
(3) provide grounded design principles for a GenAI instructional material design agent. These findings will inform the creation of AI tools that are context-aware, evidence-anchored, and pedagogically aligned.

\section{Experiments}

\begin{figure*}[htbp]
    \centering
    \includegraphics[width=1 \textwidth]{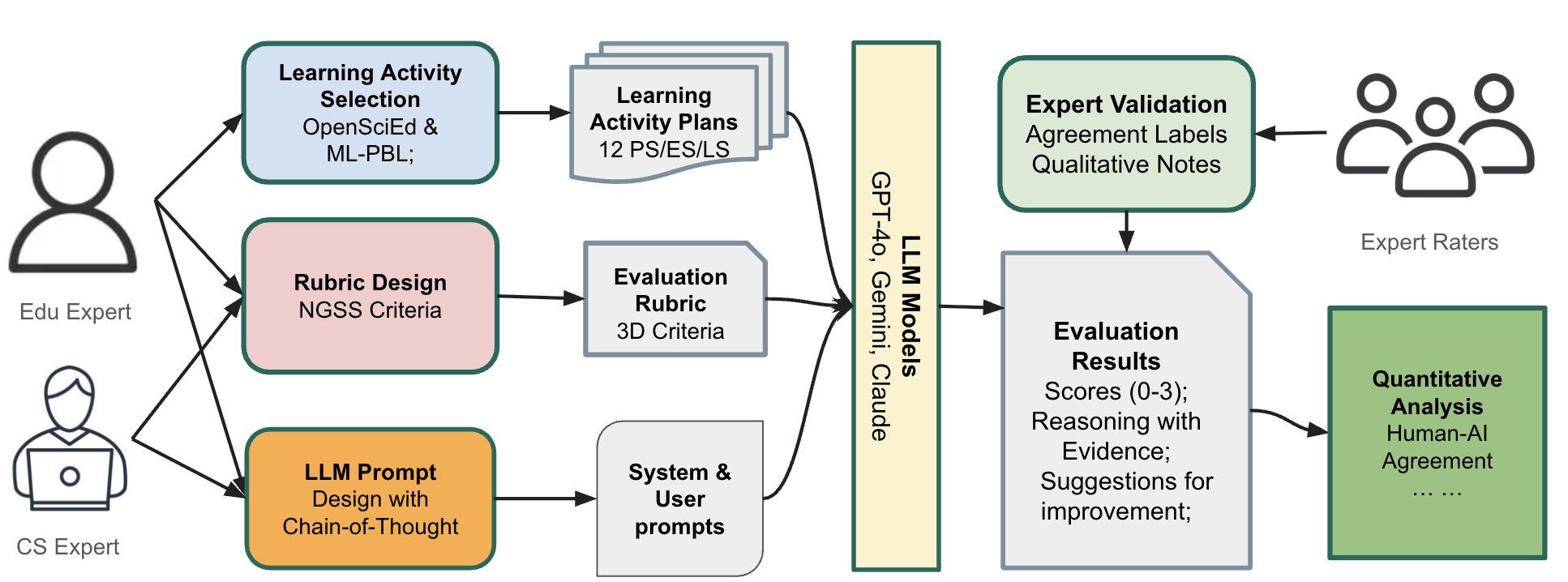}
    \caption{System workflow for evaluating learning activities with LLMs. 
    }
    \label{fig:workflow}
\end{figure*}

Our study was designed to systematically compare the evaluation outputs of multiple LLMs against human expert judgments when applying the EQuIP rubric to high-quality K--12 science instructional materials. We selected 12 validated science curriculum units spanning life sciences (LS), physical sciences (PS), and Earth sciences (ESS). These units were sourced from \emph{OpenSciEd} and \emph{Multiple Literacies in Project-Based Learning}, both recognized for their NGSS alignment and instructional quality. Figure~\ref{fig:workflow} presents the research design and workflow overview.

\subsection{Experiment setup} 
\noindent
Three state-of-the-art LLMs were selected for evaluation: 1.~GPT-4o~\cite{openai2024gpt4o}, 2.~Gemini 2.5 Pro~\cite{google2024gemini}, 2.~Claude Sonnet 4~\cite{anthropic2024claude}.
\noindent
We used the 9-item \emph{Educators Evaluating the Quality of Instructional Products} (EQuIP) rubric \cite{achieve2016equip} as the evaluation framework. Each rubric item was scored on a 0--3 scale, with written rationales required for each score. 
\noindent
For each combination of \{model, class activity document, rubric item\}, we prompted the LLM use of EQuIP criteria to: \emph{First}, assign a numerical score (0--3). \emph{Second}, provide a written rationale that justifies the score. \emph{Finally}, give a suggestion for improvement based on the rationale evidence. 
This resulted in $12 \times 3 \times 9 = 648$ outputs (numerical score + rationale + suggestion).  
Two experienced science education experts independently reviewed all outputs, marking agreement (1) or disagreement (0) for both scores and rationales, and providing qualitative reflections on the AI reasoning. The example output for one learning activity case is shown in Table \ref{tab:model-human-example}.

\paragraph{Evaluation Metrics} After that, we computed: Agreement rate (\%) between experts and each model for scores and rationales, and the Inter-rater reliability between human experts (Cohen’s $\kappa$, Fleiss’ $\kappa$) for reliability.

\paragraph{Prompt Design}

As shown in Table~\ref{tab:prompts}, we adopted a standardized two-part prompt structure for all models to ensure consistency across evaluations. The first component was a \emph{system prompt}, which explicitly established the model’s role as a professional school teacher and framed the evaluation task in terms of the NGSS-aligned rubric criteria. The second component was a \emph{user prompt}, which supplied the specific curriculum activity along with detailed evaluation instructions for each rubric dimension. This evaluation required the model to: (1) assign an \emph{Evidence of Quality} score (0–3), (2) justify its score through \emph{reasoning supported by concrete evidence from the activity}, and (3) propose \emph{suggestions for improvement}. 

Importantly, this prompt design was co-developed by education researchers and computer science researchers to balance pedagogical authenticity with computational rigor. This interdisciplinary collaboration ensured that the evaluation task mirrored authentic teacher-led rubric practices while incorporating best practices in LLM prompting. Specifically, the design integrates chain-of-thought (CoT) prompting \cite{wei2022chain}, requiring models to articulate reasoning with evidence before assigning scores and suggesting improvements. In this way, CoT functions not only as a mechanism for eliciting step-by-step reasoning but also as an educationally authentic workflow that reflects real-world teacher judgment, thereby enhancing both ecological validity and computational robustness.  


\begin{sidewaystable*}[htbp]
\centering
\small
\setlength{\tabcolsep}{3pt} 
\caption{Example of model output and human agreement for one K\textendash 12 lesson activity. Human Agreement is coded as 1 (expert agrees with the model) or 0 (expert disagrees).}
\label{tab:model-human-example}
\begin{tabular}{p{0.14\linewidth} p{0.09\linewidth} p{0.06\linewidth} p{0.29\linewidth} p{0.06\linewidth} p{0.29\linewidth} p{0.07\linewidth}}
\toprule
\textbf{Criterion} &
\textbf{Quality (0--3)} &
\textbf{Human Agree.} &
\textbf{Evidence \& Reasoning} &
\textbf{Human Agree.} &
\textbf{Suggestions for Improvement} &
\textbf{Human Agree.} \\
\midrule
I.A: Explaining Phenomena / Designing Solutions &
3 (Extensive) &
1 &
Students explore balancing sculptures by defining success criteria, trying designs, revising them, and connecting ideas to artistic and physical science principles (e.g., symmetry, balance point). Shown in lesson steps: building symmetrical sculptures, revising ideas, referencing ``A Successful Sculpture Will'' chart. &
1 &
Could further encourage students to generate their own driving questions tied to personal experiences with balance in playground or home structures, for even richer sense-making. &
1 \\
I.B: Three Dimensions &
3 (Extensive) &
1 &
SEPs: Ask questions, plan and carry out investigations (e.g., testing sculpture stability); DCIs: PS2.A and ETS1-1 on forces and design constraints; CCCs: cause and effect of symmetry/asymmetry on balance. Integrated throughout build-and-revise cycles. &
1 &
Could deepen explicit links to crosscutting concepts by having students track cause-and-effect patterns on a class chart over multiple lessons. &
0 \\
I.C: Integrating Three Dimensions &
3 (Extensive) &
1 &
Students integrate practices (investigation), core ideas (forces/balance), and CCCs (cause and effect) during design, measurement, and revision phases (p.\ 8--9, p.\ 15--19). &
1 &
Could highlight integration by asking students to reflect specifically on how SEPs, DCIs, and CCCs came together at lesson close. &
0 \\
II.A: Relevance and Authenticity &
2 (Adequate) &
1 &
Students experience authentic design challenges, build sculptures using realistic materials, and read about real-world art/science connections. Lesson references connections to local art or culture (p.\ 5 Preparation Checklist). &
1 &
Expand explicit suggestions on linking to community sculptures or local artists, e.g., invite photos of public sculptures from students’ neighborhoods. &
1 \\
II.B: Student Ideas &
3 (Extensive) &
1 &
Students repeatedly express, record, justify, and revise their ideas through charts, discussions, handouts, and partner talks (p.\ 13, 24--27). &
1 &
Encourage structured peer feedback routines (e.g., ``two stars and a wish'') to strengthen discussion quality. &
1 \\
II.C: Scientific Accuracy &
3 (Extensive) &
1 &
Scientifically accurate content on balance, symmetry, forces, and design principles. Vocabulary such as ``balance point'' is scaffolded and precisely defined. &
1 &
No major improvements; could preview tie-in to later force lessons for coherence. &
1 \\
III.A: Monitoring 3D Performances &
3 (Extensive) &
1 &
Lesson includes formative opportunities with pre-assessment (criteria and constraints chart) and Building Understandings discussions (p.\ 10, 24). &
1 &
Could add a final exit ticket to check 3D integration at closure. &
1 \\
III.B: Feedback &
3 (Extensive) &
1 &
Teachers use prompts, discussion checks, and Building Understandings to give feedback during sculpture building (p.\ 19--20, 26--28). &
1 &
Could add peer assessment checklists for sculpture stability. &
1 \\
III.C: Unbiased Tasks/Items &
3 (Extensive) &
1 &
Materials and instructions are accessible to diverse learners, with strong universal design features (visual supports, language prompts, varied roles) (p.\ 16--19). &
1 &
Could check for culturally relevant references to validate diverse backgrounds. &
1 \\
\bottomrule
\end{tabular}
\end{sidewaystable*}

\section{Results}

\paragraph{Descriptive statistics of score}
We computed descriptive statistics of mean scores by dimension, topic, and model (Table~\ref{tab:score-means}). 
Overall, models average close to 2.66 on the 0--3 scale. By topic, Earth Science had slightly higher average scores (2.69), followed by Life Science (2.67) and Physical Science (2.64). 
By model, Gemini gave the highest average scores (2.96), GPT was close (2.81), while Claude was much stricter (2.18). 
This confirms the patterns seen in the pairwise agreement analysis: Gemini and GPT are generous and closely aligned, while Claude is stricter, producing lower scores and less overlap with the other two models.

\begin{table}[H]
\centering

\caption{Mean score across dimensions, topics, and models.}
\label{tab:score-means}
\begin{tabular}{lcc}
\toprule
\textbf{Group} & \textbf{Category} & \textbf{Score Mean} \\
\midrule
Overall & All Dimensions & 2.664 \\
\midrule
Topic & Earth Science (ES) & 2.686 \\
      & Life Science (LS)  & 2.667 \\
      & Physical Science (PS) & 2.639 \\
\midrule
Model & OpenAI GPT-4o    & 2.806 \\
      & Claude Sonnet 4 & 2.182 \\
      & Gemini 2.5 Pro & 2.955 \\
      
\bottomrule
\end{tabular}

\end{table}
\vspace{-1em}

\subsubsection{Human-AI Agreement} 
\begin{figure*}[t]
\centering
\includegraphics[width=\textwidth]{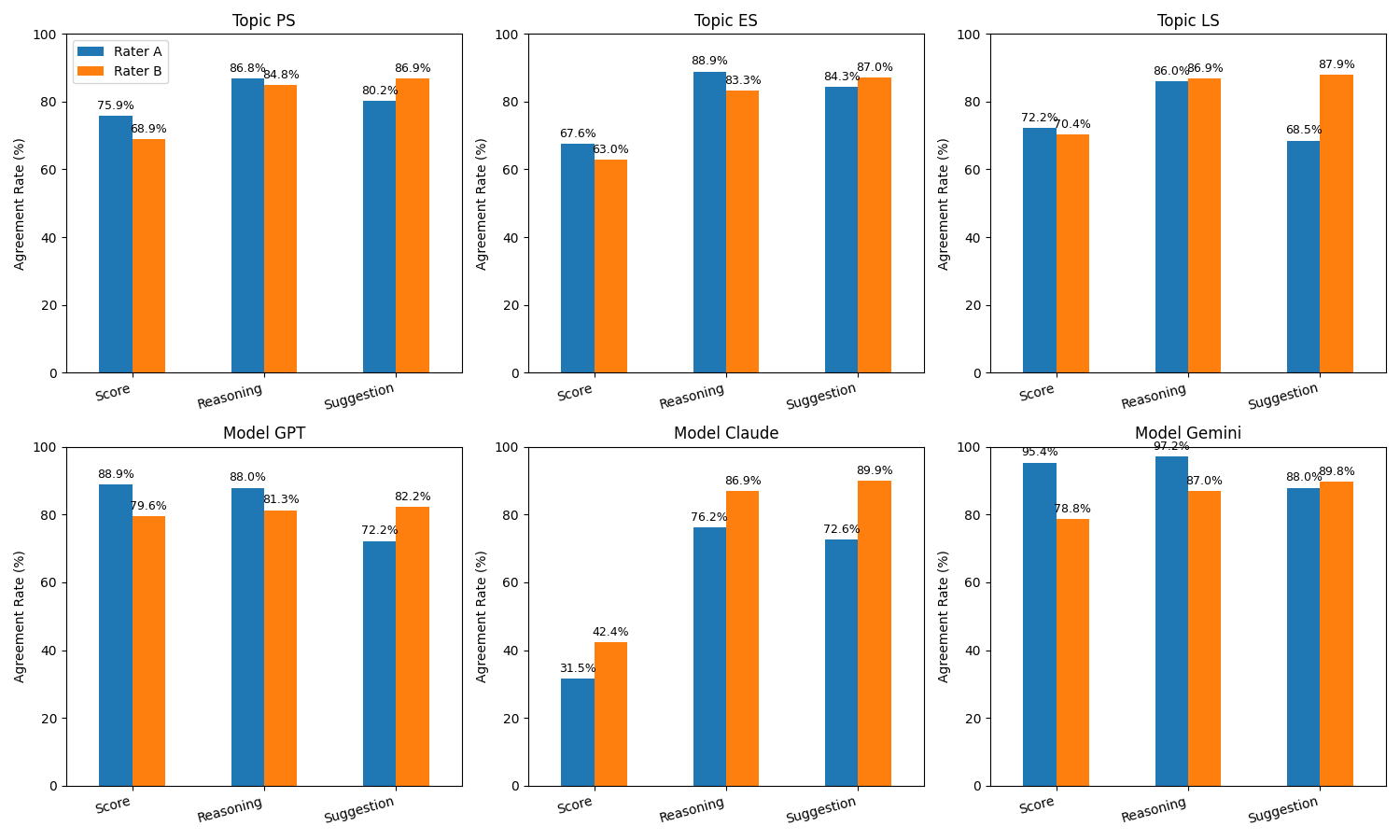} 
\caption{Human--LLM agreement rates (\%) by topic (top row) and by model (bottom row). Each bar shows Rater A (blue) and Rater B (orange).}
\label{fig:agreements}
\end{figure*}

To evaluate how closely human raters align with LLMs, we quantify alignment as the percentage of rubric instances where each human rater (A, B) agrees with an LLM's output (score, rationale, or suggestion). Table~\ref{tab:overall-align} summarizes overall agreement across all units and models, Figure~\ref{fig:agreements} visualizes agreement by topic and by model, and Table~\ref{tab:model-align} aggregates agreement per model.
\vspace{-0.5em}
\begin{table}[H]
\centering
\small
\caption{Overall Human--LLM agreement rates (\%).}
\label{tab:overall-align}
\begin{tabular}{lccc}
\toprule
\textbf{Group} & \textbf{Score} & \textbf{Reasoning} & \textbf{Suggestion} \\
\midrule
Rater A & 71.9 & 87.2 & 77.6 \\
Rater B & 67.3 & 85.0 & 87.3 \\
\midrule
\textbf{Mean (A,B)} & \textbf{69.6} & \textbf{86.1} & \textbf{82.5} \\
\bottomrule
\end{tabular}

\end{table}
\vspace{-1em}
Humans show a decent average alignment with LLMs on scores (mean 69.6\%). Agreement is notably higher on reasoning (86.1\%) and suggestions (82.5\%), indicating raters found more to agree with reasoning and improvement content than in the raw numeric ratings. 

\paragraph{Agreement by LLM Models}

Table~\ref{tab:model-align} combines Rater A and B by simple averaging for each model. Gemini yields the strongest score alignment on average (87.1\%) and also leads on reasoning (92.1\%) and suggestions (88.9\%).
GPT is close behind on scores (84.3\%) and \emph{reasoning} (84.7\%).
Claude shows very low score alignment (37.0\%) but comparatively good reasoning (81.6\%) and suggestions (81.3\%). That is, raters largely disliked the numeric scores but liked the suggestions (\(+44.3\) percentage points higher on suggestions than scores).

\vspace{-0.5em}
\begin{table}[H]
\centering
\caption{Mean agreement (\%) by model; averages combine Rater A and B.}
\label{tab:model-align}
\begin{tabular}{lccc}
\toprule
\textbf{Model} & \textbf{Score} & \textbf{Reasoning} & \textbf{Suggestion} \\
\midrule
GPT & 84.3 & 84.7 & 77.2 \\
Claude & 37.0 & 81.6 & 81.3 \\
Gemini & \textbf{87.1} & \textbf{92.1} & \textbf{88.9} \\
\bottomrule
\end{tabular}

\end{table}
\vspace{-1em}

\paragraph{Agreement by Science Domain}
Figure~\ref{fig:agreements} shows that Earth Sciences shows the lowest score agreement (A: 67.6\%, B: 63.0\%; mean \(\approx 65\%\)), suggesting scoring is harder for that domain. Physical Sciences (A: 75.9\%, B: 68.9\%; mean \(\approx 72\%\)) and Life Sciences (A: 72.2\%, B: 70.4\%; mean \(\approx 71\%\)) have moderately higher score agreement.
Across all three domains, agreement on reasoning and suggestions remains higher than on scores (e.g., Earth Science reasoning mean \(\approx 86\%\), suggestions mean \(\approx 86\%\)).

\subsubsection{Inter-rater reliability}
\begin{figure}[H]
\centering
\includegraphics[width=0.75\textwidth]{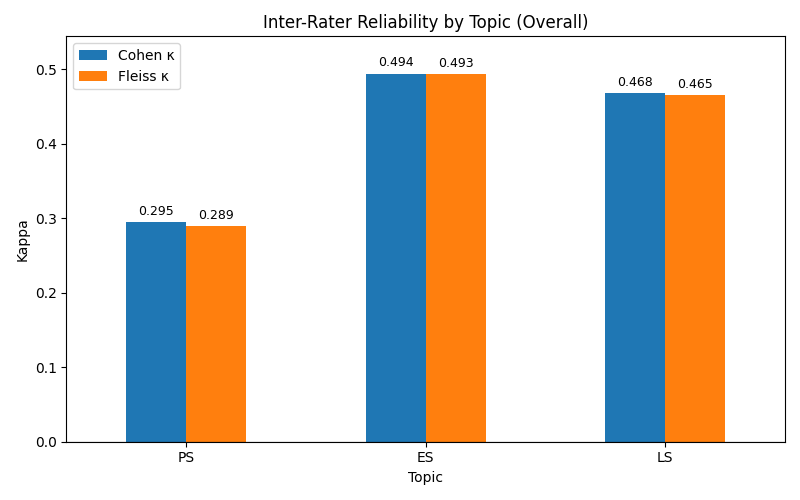}
\caption{Inter-rater reliability (Cohen’s $\kappa$ and Fleiss’ $\kappa$) by topic.}
\label{fig:kappa}
\end{figure}

We also examined the reliability between the two human raters using both Cohen’s $\kappa$ and Fleiss’ $\kappa$ across the three science domains. 
Figure~\ref{fig:kappa} shows that reliability was lowest in Physical Science ($\kappa \approx 0.29$), while Earth Science ($\kappa \approx 0.49$) and Life Science ($\kappa \approx 0.47$) showed moderate agreement. 
This suggests that judging Physical Science activities may be more subjective or that the rubric is harder to apply consistently in this domain.

In short: (i) numeric \emph{scores} attract more disagreement than textual \emph{reasoning} and \emph{suggestions}; (ii) Gemini achieves the strongest overall alignment, GPT is close, and Claude’s numeric scores underperform while its suggestions resonate with human raters; and (iii) Earth Science scoring is comparatively difficult, with the lowest human–LLM alignment among the three domains.

\subsubsection{Agreement among LLMs}
\begin{figure}[t]
\centering
\includegraphics[width=0.75\textwidth]{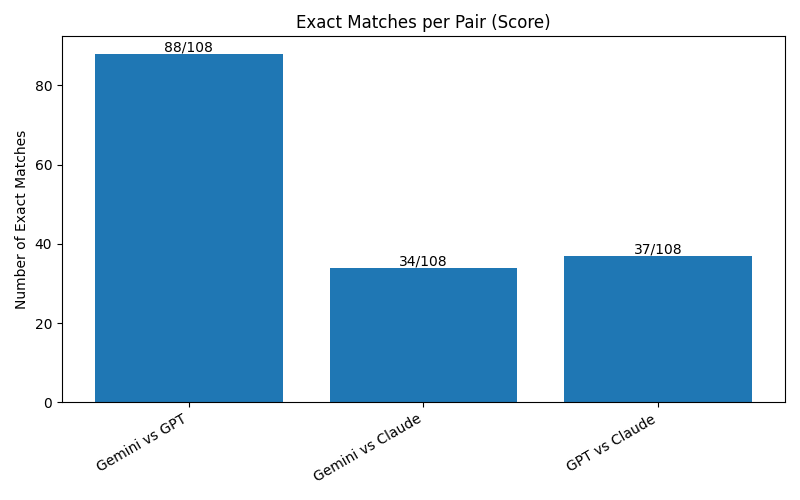}
\caption{Pairwise exact matches of scores between LLMs.}
\label{fig:pairwise}
\end{figure}

\begin{figure}[t]
\centering
\includegraphics[width=0.75\textwidth]{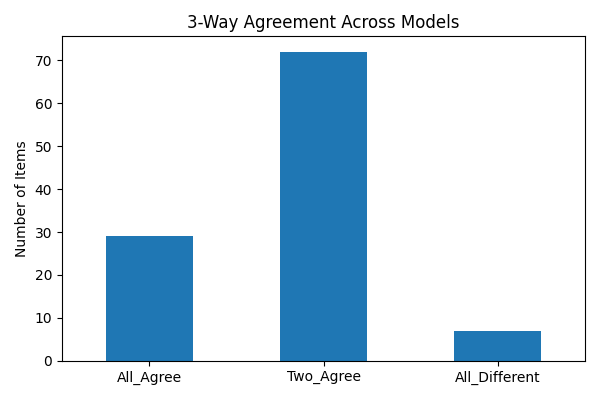}
\caption{Three-way agreement across models.}
\label{fig:threeway}
\end{figure}

We also calculated agreement between the three LLMs. 
As shown in Figure~\ref{fig:pairwise}, Gemini and GPT share the largest overlap in their scores (88/108 exact matches), while Claude aligns much less frequently with either Gemini (34/108) or GPT (37/108). 
Claude tends to assign lower scores overall, consistent with the lower mean score statistics reported below. 
Nevertheless, when examining three-way agreement (Figure~\ref{fig:threeway}), most cases fall into the \textit{Two\_Agree} category, meaning that two models agreed and one diverged. 
  
The possible explanation of the difference in performance patterns is that GPT-4o and Gemini are both built as multimodal foundation models with broad generalization capabilities. GPT-4o is trained end-to-end on text, image, audio, and video in a unified Transformer architecture and human feedback \cite{attentionyouneed}, allowing it to integrate information across modalities. Gemini, likewise, is a multimodal family of models designed to handle diverse inputs (text, audio, image, video) with interleaved context and high generality. These architectural and training similarities likely contribute to their higher overlap in scoring (88/108 exact matches).  
In contrast, Claude emphasizes \emph{safety alignment} and conservative generation. It is trained via Constitutional AI, a reinforcement learning-based method that strongly favors "helpful, honest, and harmless" outputs \cite{anthropic2025claude_systemcard}. These constraints likely bias Claude toward more conservative scoring, reducing overlap with other models.

GPT-4o and Gemini behave more like holistic scorers, awarding partial credit and recognizing broader patterns of student reasoning—thereby aligning more closely with human raters. Claude, by contrast, resembles a strict grader, penalizing deviations more heavily and emphasizing precise correctness. This distinction explains its consistently lower mean scores and lower exact-match agreement, despite still aligning on reasoning and suggestions.

\paragraph{Implementation.}
Both prompts were presented to the LLM alongside:
1. The full text of the NGSS 3D Learning Activities evaluation criteria (attached as reference material).
2.  The complete lesson description for the unit being evaluated.

Tables~\ref{tab:prompts}(above) illustrate the prompt templates adapted for numerical scoring and rationale generation.  
For this task, the “score” and “rationale” were combined in a single tabular output per rubric category.

\subsection{Case-Based Insights: Divergent Interpretations of Quality}

Our quantitative findings revealed two key trends. First, \textbf{inter-rater reliability among human experts was only moderate}, with especially low agreement in Physical Sciences tasks ($\kappa \approx 0.29$). \textbf{Second, LLMs differed significantly in scoring behavior}: \emph{Gemini assigned the highest average scores (2.96), followed by GPT (2.81), while Claude was notably more stringent (2.18)}. Interestingly, although models often disagreed numerically, their rationales for scoring were frequently \emph{aligned}—suggesting that while they recognized similar instructional features, they weighted them differently.

To make these patterns more tangible, we present two illustrative cases: one spotlighting human disagreement, and the other highlighting AI model divergence.

\paragraph{Human Disagreement: Diverging Views on Integration in an Open-Ended Task}

This case exemplifies the first trend---\textit{higher disagreement among human raters} on exploratory, open-ended lessons. The lesson in question comes from \textbf{OpenSciEd Unit 3.1, Lesson 2}, designed for Grade 3 Physical Sciences. In this activity, students are invited to build sculptures that balance on a point using everyday classroom materials. The prompt encourages them to explore: \textit{"Can you build a sculpture that balances? What makes it stable or unstable?"}

\textbf{Rater A} gave a score of \textbf{3}, highlighting how the task "encourages iterative sensemaking, with students predicting, testing, and revising their models." He noted that the lesson taps into the concept of structure and function, and emphasized that students are supported in generating their own explanations, a hallmark of three-dimensional learning.

\textbf{Rater B}, however, gave the same lesson a score of \textbf{1}, reasoning that "students are not explicitly prompted to use disciplinary core ideas like force or balance," and warning that the task could be interpreted as "more of an arts-and-crafts exercise than a science investigation."

This case reveals a profound \textbf{epistemic divide}: Rater A privileges emergent, embodied reasoning as valid science learning, while Rater B demands explicit content prompts and conceptual alignment. The disagreement mirrors our quantitative reliability results, where Physical Sciences showed the lowest inter-rater agreement. In classrooms, this could mean that how a task is judged may depend heavily on the Judge's underlying beliefs about what counts as "real" science and their epistemological investigation approaches.

\paragraph{Model Disagreement: Contrasting AI Epistemologies in a Thermal Insulation Task}

The second case showcases the second trend---\textit{divergence in scoring among models}. This activity is taken from \textbf{OpenSciEd Unit L4.1, Lesson 4} for Grade 5 students, focusing on Physical Sciences. Students are tasked with designing and testing insulation materials to keep water warm, using temperature data to iterate on their designs. The prompt asks: \textit{"How can we keep the water in the cup from getting cold? Build a prototype that minimizes heat loss."}

\textbf{Claude} scored the lesson a \textbf{1}, criticizing that "students were not explicitly required to articulate why their design worked," and that the task "lacked mandatory SEP or CCC alignment." Claude's explanation revealed a rule-based view of instructional quality---one that values precision, structure, and explicit standards alignment.

\textbf{GPT} gave the lesson a \textbf{2}, describing the activity as "a real-world application of engineering practices" but without offering specific evidence or elaboration. Its middle-ground stance appeared cautious, leaning on safe generalities rather than analytical depth.

\textbf{Gemini}, in contrast, assigned a score of \textbf{3}, quoting directly from the lesson and supporting the opportunity for iterative design and use of empirical data. Gemini emphasized that "students analyze real temperature data and revise their models," framing the task as one rich with implicit opportunities for scientific reasoning and knowledge construction.

These differences illuminate \textbf{three distinct AI epistemologies}: \textit{Claude as prescriptive and strict}, \textit{GPT as neutral and shallow}, and \textit{Gemini as interpretive and holistic}. Notably, these narrative patterns echo our quantitative findings: Claude was the strictest model, GPT remained moderate, and Gemini aligned most closely with expert rationales.

\paragraph{Design Implications}

These two cases challenge the assumption that disagreement in scoring is mere noise. Instead, they reflect deeply held beliefs---human or machine---about what counts as good instruction. For designers of educational AI, this means that educational AI systems should not be optimized for consensus alone. Rather, they could be designed to surface multiple justifications, each revealing a different lens through which to view instructional quality. Such transparency can invite teacher reflection, support professional learning, and ultimately contribute to more nuanced and equitable classroom assessment.

\section{Discussions}
Our findings offer critical insights for designing trustworthy, human-centered AI systems
in education—particularly those that evaluate or co-develop instructional materials.
Rather than optimizing solely for scoring accuracy, such systems must attend to the
interpretive and epistemic dimensions of instructional judgment, a longstanding concern
in educational assessment research \cite{Li2024CognitiveSynergy}.

First, quantitative metrics like score agreement and rubric alignment provide useful
diagnostics, but they obscure the why behind agreement or disagreement. Prior work
has shown that agreement coefficients can mask substantial differences in underlying
judgment criteria and reasoning processes for both human and AI raters
\cite{landis1977measurement,hackl2023gpt,tate2024can}. Our case analyses similarly reveal that similar
scores can conceal deep divergences in pedagogical interpretation, while differing scores
may still reflect partially aligned reasoning. These tensions call for AI systems that
prioritize epistemic transparency—surfacing the interpretive logics that shape both
human and model evaluations rather than presenting scores as self-explanatory outcomes.

Second, the distinct behavior of LLMs—Claude's rule-based precision, GPT's cautious
neutrality, and Gemini's holistic sensemaking—highlights that no single model captures the
full range of pedagogical values held by human experts. Recent studies of AI-assisted
evaluation and content generation likewise report substantial variation across models in
how instructional quality is operationalized and justified
\cite{huang2024application,AbdulWahid2025CurriculumAlignedMCQ,Clark2025AutoEvaluation}. We therefore advocate for a shift toward
model multiplicity, enabling users to compare diverse rationales and toggle
between different epistemic stances. Such a perspective-surfacing approach aligns with
emerging frameworks that position AI as a partner in curriculum development rather than a
replacement for professional judgment \cite{Dickey2023GAIDE,Tavakoli2021Hybrid}.

Third, human disagreement itself should not be treated as noise. Decades of assessment
research have established that rater disagreement often reflects legitimate differences
in interpretive frames rather than error \cite{cohen1960coefficient,fleiss1971measuring}. In this study,
disagreement provided a valuable signal for identifying sites of pedagogical
uncertainty—where multiple interpretations of instructional quality are valid and
context-dependent. These moments frequently reflect deeper dilemmas, such as balancing
emergent reasoning with conceptual rigor or honoring cultural relevance alongside
disciplinary alignment, echoing persistent challenges in NGSS-aligned curriculum design
\cite{NRC2012,NASEM2018}.

Technically, designing such disagreement-aware systems requires moving beyond single-label
optimization. Prior evaluations of AI raters demonstrate that consistency and accuracy
alone are insufficient indicators of educational usefulness
\cite{hackl2023gpt,tate2024can}. Future models might incorporate disagreement-aware loss
functions, multi-rater calibration modules, or uncertainty estimates to represent both
consensus and epistemic diversity. At scale, this could enable adaptive AI personas
aligned with different teaching philosophies and foster richer human--AI collaboration.

Looking ahead, we are building a multi-agent AI high-quality curriculum development system
designed to operationalize these insights—allowing educators to view diverse model
rationales, explore tensions, and co-interpret instructional quality. This direction is
consistent with calls for human-centered AI systems that support teacher agency and
professional sensemaking rather than automate instructional decision-making
\cite{holstein2019co,zawacki2019systematic,Jaramillo2024AIDriven,AbdulWahid2025CurriculumAlignedMCQ}.

In sum, this study contributes to a broader vision of educational AI: one that augments
rather than automates, collaborates rather than dictates, and learns with—not instead of—
teachers. Recognizing instructional judgment as a deeply interpretive and socially situated
practice is essential for developing AI tools that are pedagogically principled,
context-aware, and ethically grounded.

\subsection*{Contributions and Implications}

This study contributes to both theoretical understanding and practical design of
AI-supported curriculum evaluation. Theoretically, it reframes human--AI agreement
not as a binary indicator of accuracy but as an epistemic signal that reflects
distinct interpretive orientations embedded in both expert and model judgments.
This perspective extends prior human-in-the-loop validation research by treating
disagreement as a resource for analysis rather than noise to be minimized.

Practically, our findings inform the development of more reliable and context-aware
GenAI evaluation systems for K--12 instructional materials. The observed divergence
between numeric rubric scores and qualitative rationales suggests that future AI
tools should prioritize explanatory transparency and comparative reasoning over
single-score outputs. Moreover, the model-specific disagreement patterns identified
in this study offer concrete heuristics for configuring multi-agent or ensemble-based
GenAI systems that support educator sensemaking and professional judgment.

\paragraph{Limitations and Future Directions.}
Several limitations should be acknowledged. First, this study examined a relatively
small set of 12 high-quality science curriculum units drawn from established
instructional programs, which may limit generalizability to locally developed or
lower-quality materials. Second, expert validation was conducted by a small number
of science education specialists; while their disagreements were analytically
productive, additional expert communities may surface alternative interpretive
perspectives. Third, the analysis focused on rubric-based evaluation rather than
classroom enactment, leaving open questions about how AI-supported evaluations
translate into teachers' instructional decision-making in practice.

Future work will extend this framework to larger and more diverse curriculum
collections, incorporate broader expert participation, and address practical
considerations such as scalability, computational cost, and educator adoption of
disagreement-aware AI evaluation tools.

\section{Conclusion}
This paper examined how human experts validate and interpret AI-generated evaluations
of high-quality K--12 science instructional materials across multiple large language
models. By combining comparative multi-LLM analysis with expert judgment, we showed
that patterns of agreement and disagreement reflect underlying epistemic orientations
rather than simple model errors.

By treating disagreement as an analytic signal rather than a failure mode, this study
advances a human-centered perspective on AI-supported curriculum evaluation. The
design implications derived from expert reasoning highlight the need for
disagreement-aware, explanatory, and context-sensitive GenAI systems that support
teacher sensemaking rather than automate instructional judgment. Together, these
findings contribute to ongoing efforts to develop trustworthy AI systems grounded in
educational values and professional practice.

\bibliographystyle{splncs03}
\bibliography{aaai2026}
\end{document}